\documentclass{elsart}

 \usepackage{graphicx}
\begin{document}

\begin{frontmatter}



\title{Monte Carlo studies of a Finsler geometric surface model}


\author{Hiroshi Koibuchi}
\ead{koibuchi@mech.ibaraki-ct.ac.jp}

\address{Department of Mechanical and Systems Engineering, Ibaraki National College of Technology, 
Nakane 866, Hitachinaka,  Ibaraki 312-8508, Japan}

\author{Hideo Sekino}

\address{Computer Science and Engineerings, Toyohashi University of Technology, 
Hibarigaoka 1-1, Tenpakuchou, Toyohashi, Aichi 441-8580, Japan}

\begin{abstract}
This paper presents a new type of surface models constructed on the basis of Finsler geometry. A Finsler metric is defined on the surface by using an underlying vector field, which is an in-plane tilt order. According to the orientation of the vector field, the Finsler length becomes dependent on both position and direction on the surface, and for this reason the parameters such as the surface tension and bending rigidity become anisotropic. To confirm that the model is well-defined, we perform Monte Carlo simulations under several isotropic conditions such as those given by random vector fields. The results are comparable to those of previous simulations of the conventional model. It is also found that a tubular phase appears when the vector field is constant. Moreover, we find that the tilts form the Kosterlitz-Thouless and low temperature configurations, which correspond to two different anisotropic phases such as disk and tubular, in the model in which the tilt variable is assumed to be a dynamical variable. This confirms that the model in this paper may be used as an anisotropic model for membranes.        
\end{abstract}

\end{frontmatter}

% main text
%\section{}
%\label{}
%----------------------------------------------------------
\section{Introduction}\label{intro}
%----------------------------------------------------------
A membrane is understood as a mapping from two-dimensional surface $M$ to ${\bf R}^3$ \cite{FDAVID-SMMS2004}. If $M$ is of sphere topology, the image $X(M)$ corresponds to a spherical membrane. The shape of $X(M)$ is governed by surface tension energy and bending energy, as it is assumed in the surface model of Helfrich and Polyakov (HP) \cite{HELFRICH-1973,POLYAKOV-NPB1986}. As a general framework for phase transitions, the Landau-Ginzburg free energy is also assumed \cite{Radzihovsky-SMMS2004,WIESE-PTCP19-2000,Bowick-PREP2001} for membranes, where the tangential vector $\partial X$ of the surface is an order parameter.  Owing to such mathematical backgrounds, the surface shape and its phase structure have been extensively studied \cite{P-L-1985PRL,DavidGuitter-1988EPL,KLEINERT-PLB1986,KANTOR-KARDAR-NELSON-PRL1986,Kardar-Nelson-PRA1988}. 

Almost all shape transformations in membranes are concerned with anisotropic phases such as tubular and planar (or prolate, oblate) phases. An anisotropic phase was predicted in a surface model with a bending rigidity which is anisotropic in the internal direction of the surface \cite{Radzihovsky-Toner-PRL1995,Radzihovsky-Toner-PRE1998}, and the existence of such anisotropic surface was numerically confirmed \cite{BOWICK-etal-PRL1997}. These anisotropic surface models have also been studied by non-perturbative renormalization group formalization \cite{Essafi-Kownack-Mouhanna-2011PRL}. Moreover, a tubular surface can also be seen in a surface model with elastic skeleton, where  one-dimensional bending energy is assumed \cite{Koibuchi-PRE2007}. 

However, the origin of the anisotropy in membranes still remains to be clarified. Indeed, it is unclear why the bending rigidity becomes isotropic (or anisotropic) \cite{Radzihovsky-Toner-PRL1995,Radzihovsky-Toner-PRE1998}. Furthermore, it is well-known that there exist anisotropic membranes without skeletons \cite{Seifert-EPL1996}.  

An origin of the anisotropic surface shape is considered to be connected with the direction of liquid crystal molecules in liquid crystal elastomers (LCEs) membrane \cite{Lubensky-etal-PRE2003,Xing-Radzihovsky-ANPHYS2008}. The non-polar orientation property is always assumed for the liquid crystal molecules. For this reason the mechanical strength of LCEs depends on whether the molecules are aligned or not. On the other hand, the dynamical variables of the HP surface model mentioned above are the surface position $X$ and the metric $g$ of $M$. For this reason, one possible explanation for the anisotropic surface shape is that the metric $g$ is anisotropic and this anisotropy is represented in $X(M)$ in the HP model. Thus, recalling that the Finsler metric reflects a space anisotropy in general \cite{Matsumoto-SKB1975}, we can implement the molecular orientation property in the HP model by replacing the Riemmanian metric $g$ with the Finsler metric $g^F$. In other words, the molecular orientation property can geometrically be implemented in the context of HP model on the basis of Finsler geometry, and as a consequence the surface anisotropy can be explained naturally.   

Therefore, it is interesting to study a surface model on the basis of Finsler geometry \cite{Matsumoto-SKB1975}. In Finsler geometry, an infinitesimal length unit is considered to be dependent on the directions, and it gives a more general framework than the Riemannian geometry \cite{Bogoslovsky-Goenner-PLA1998,OtsukaTanaka-PLA2010}. Because of the length unit anisotropy, we expect that the surface force in membranes (such as the surface tension and the bending rigidity) naturally becomes anisotropic if the Finsler geometry can be implemented in the surface models. Those Finsler geometric (FG) surface models are expected to provide us a natural framework for describing anisotropic shape transformations in membranes.

In this paper, a Finsler geometric surface model for membranes is studied and Monte Carlo (MC) simulation data are presented. This model is constructed by extending a discrete surface model of Helfrich and Polyakov such that the metric function is replaced by a Finsler metric. The assumed Finsler metric is defined by using a vector filed ${\bf v}$ on $M$. Since the ${\bf v}$ has its own direction on the surface, this vector field ${\bf v}$ is considered to be an origin of surface anisotropy. We should note that the isotropy is restored if ${\bf v}$ is given locally at random. In this case, the FG model should have the same phase structure as the conventional surface model of Helfrich and Polyakov. We firstly check this under several conditions and confirm that the FG surface model is well-defined. Nextly, it is demonstrated that anisotropic surfaces are obtained when ${\bf v}$ is constant and treated as a dynamical variable with the Heisenberg spin model Hamiltonian.

%----------------------------------------------------------
\section{Finsler geometric surface model}\label{model}
%----------------------------------------------------------
%----------------------------------------------------------
\subsection{Elements of Finsler geometry}\label{Finsler_geometry}
%----------------------------------------------------------
In this subsection, we briefly summarize the elements of Finsler geometry \cite{Matsumoto-SKB1975}. Let $M$ be a  two-dimensional manifold, and let $C$ be a curve on $M$ such that $ C\ni t\mapsto x(t)\in M$. We call $M$  a {\it Finsler space} if there exists a Finsler funtion $L$ on $M$ such that the Finsler length $s$ of the curve $C$ is given by
\begin{equation}
\label{length_curve}
s=\int_{t_0}^t L(x,y)dt,
\end{equation}
where $L(x,y)$ is a homogeneous function of degree $1$ with respect to $y$. The symbols $x\!=\!(x_1,x_2)$ and $y\!=\!(y_1,y_2)\!=\!(dx_1/dt,dx_2/dt)$ in $L$ denote a point on $C$ and a tangential vector at $x$ with the direction along which $t$ increases, respectively. Thus, we have 
\begin{equation}
\label{length_constraint}
L(x,ky)=kL(x,y)
\end{equation}
for any positive $k$. This equation implies that the Finsler length $s$ of the curve $C$ is independent of the parameter $t$. For this reason, the Finsler length $s$ depends only on the ratio $y_2/y_1$, because $y\!=\!(y_1,y_2)\!=\!y_1(1,y_2/y_1)$ can be replaced by  $(1,y_2/y_1)$ in Eq. (\ref{length_constraint}). We should note that the definition of Eq. (\ref{length_curve}) can also be written as
\begin{equation}
\label{length_derivative}
\frac {ds} {dt} =L(x,y).
\end{equation}

An example of the Finsler function $L(x(t),y(t))$ is given by using a vector field ${\bf v}$ such that
\begin{equation}
\label{example}
L(x(t),y(t))=\sqrt{\sum_i y_i^2}/\vert{\bf v}\vert,
\end{equation}
where $|{\bf v}|=\sqrt{\sum_i(dx_i/ds)^2}$. We call $\sqrt{\sum_i y_i^2}$ and $|{\bf v}|$ the Euclidean lengths. 
Note that the reparametrization  $dx_i/dt\!=\!(dx_i/ds)(ds/dt)$ allows us to write $\sqrt{\sum_i y_i^2}\!=\!\sqrt{\sum_i(dx_i/ds)^2} ds/dt$. Thus, we have $ds/dt\!=\!\sqrt{\sum_i y_i^2}/|{\bf v}|$. This leads to the expression $s$ in Eq. (\ref{length_curve}). The vector ${\bf v}$ along $C$ in Eq. (\ref{example}) can also be given by the $y$-direction component of a given vector field. In this example, the direction of ${\bf v}$ does not always need to be identical to that of $y$ and may be reverse to that of $y$. 

Let $T_xM$ be the tangential plane at $x$. Then, we have a loop made of all end points of the vectors $Y \in T_xM$ satisfying 
\begin{equation}
\label{indicatrix}
L(x,Y)=1.
\end{equation}
This equation is obtained by assuming $t\!=\!s$ in Eq. (\ref{length_derivative}). Thus, we have a Finsler length scale at $x$ such that $L(x,Y)\!=\!1$. In the case of Eq. (\ref{example}), the condition $L(x,Y)\!=\!1$ implies that $\sqrt{\sum_i Y_i^2}\!=\!\vert{\bf v}\vert$. Since the length $|{\bf v}|$ depends on its direction, we understand that the Euclidean length $\sqrt{\sum_i Y_i^2}$ depends on its direction on the loop $L(x,Y)\!=\!1$. To the contrary, the Finsler length of $Y$ is constant, which is 1 in the unit of $\vert{\bf v}\vert$, and hence it is independent of the direction on the loop. 

In the example in Eq. (\ref{example}), the Euclidean length of $s$ is given by $s|{\bf v}|$, which is identical to the Finsler length if $|{\bf v}|\!=\!1$, and hence Finsler length $s$ along $C$ is implicitly dependent on ${\bf v}$ along the direction of $y$. Thus, we find that the Finsler length $s$ defined at $x$ on $M$ depends both on the direction of ${\bf v}$ and on the length of ${\bf v}$. Since the length of ${\bf v}$ is dependent on $x$, the Finsler length depends on the position and the direction. 

%----------------------------------------------------------
\subsection{Finsler length on triangulated surfaces}\label{triangulated_surfaces}
%----------------------------------------------------------
We assume that $M$ is smoothly or piece-wise linearly triangulated in such a way that the Euclidean bond lengths are given. The vertices, the bonds, and the triangles are independently labeled on the triangulated surfaces by sequential numbers. Let $N, N_B$, and $N_T$ be respectively the total number of vertices, the total number of bonds and the total number of triangles. One additional assumption is that the bond $ij$, which is connecting two neighboring vertices $i$ and $j$, is labeled also by velocities or velocity magnitudes $v_{ij}$ and $v_{ji}$. We should note that $v_{ij}\!\not=\!v_{ji}$ in general and that $|v_{ij}|$ plays the role of the unit of Finsler length from $i$ to $j$.  Thus the bonds are labeled not only by a series of integers $1,2,\cdots,N_B$ but also by two series of real numbers $v_1,v_2,\cdots,v_{N_B}$ and $v_1^\prime,v_2^\prime,\cdots,v_{N_B}^\prime$.  It is also possible to assume that $v_i\!=\!v_i^\prime$.

%++++++++++++++++++++++++++++++++++
\begin{figure}[hbt]
\centering
\includegraphics[width=12cm,clip]{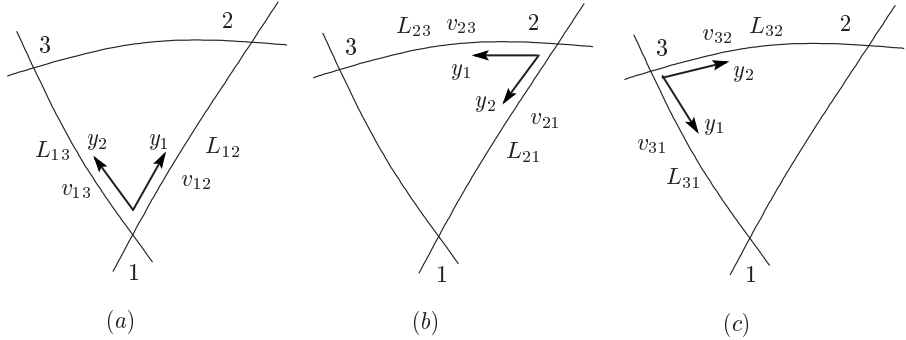}
\caption{A smooth triangle ${\it \Delta}$ in $M$ with a local coordinate (a) at the vertex $1$, (b) at the vertex $2$, and (c) at the vertex $3$. }
\label{fig-1}
\end{figure}
%++++++++++++++++++++++++++++++++++
We should comment on a natural assumption that each triangle is labeled by a single local coordinate. Indeed, we have three possible local coordinates on a triangle, and the total number of coordinates is $3N_T$ for a given triangulation of $M$. Thus, we chose a set of $N_T$ coordinates from those possible $3^{N_T}$ ones. The three possible coordinate systems of a triangle ${\it \Delta}$ are shown in Figs. \ref{fig-1}(a)--\ref{fig-1}(c). In Fig. \ref{fig-1}(a), the local coordinate of ${\it \Delta}$ is denoted by $(x_1,x_2)$ such that the origin of the coordinate axes coincides with the vertex $1$. The velocity parameters $v_{12}$ and $v_{13}$ are defined along  $y_1$ and $y_2$ respectively so that the direction of $v_{12}$ ($v_{13}$)  from $1$ to $2$ ($1$ to $3$) coincides with the direction of $y_1$ $(y_2)$. Note that the velocity $v_{21}$ is not included in the coordinate system in Fig. \ref{fig-1}(a), while it is included in Fig. \ref{fig-1}(b).

Note that the two different velocities $v_i$ and $v_i^\prime$ may be obtained from a smooth and non-constant vector field ${\bf v}$ on the surface. Indeed, suppose that ${\bf v}$ has the value only in the vertices of the triangulated surface. In such case (see Fig. \ref{fig-1}(a)), if the local coordinate origin is located at the vertex $1$, the value $v_{12}$ can be obtained from ${\bf v}(1)$ at the vertex $1$. On the contrary, $v_{21}$ is obtained from ${\bf v}(2)$ in the case of Fig. \ref{fig-1}(b). If ${\bf v}$ is not constant, then ${\bf v}(1)\!\not=\!{\bf v}(2)$, and therefore we have $v_{12}\!\not=\!v_{21}$ in general.

We firstly define a Finsler function $L(x,y)$ on ${\it \Delta}$ in Fig. \ref{fig-1}(a) such that 
\begin{equation}
\label{Finsler-function} 
L(x,y)= \left\{
       \begin{array}{@{\,}ll}
        y_1/v_{12} & \quad \quad \left({\rm on\; the\;} x_1{\rm \; axis} \right), \\
        y_2/v_{13} & \quad  \quad \left({\rm on\; the\;} x_2{\rm \; axis} \right),
       \end{array} 
       \right. \\ 
\end{equation} 
where $v_{12}\!>\!0$ and  $v_{13}\!>\!0$ are assumed. Thus the Finsler length $L_{12}$ of the bond $12$ is given by $L_{12}\!=\!L_1/v_{12}$. Indeed, $L_{12}=\int_1^2L(x,y)dt=\int_1^2(1/v_{12})y_1dt=(1/v_{12})y_1{\it \Delta}t=(1/v_{12})dx_1$. This proves that $L_{12}=L_1/v_{12}$ because $dx_1=L_1$, where $L_1$ is the Euclidean length of the bond $12$. It is also easy to see that $L_{13}=L_2/v_{13}$.
 
Secondly, we define a Regge metric \cite{REGGE-NC1961,HAMBER-LH1986,FDAVID-LH1992} on ${\it \Delta}$ in Fig. \ref{fig-1}(a) such that 
\begin{equation}
\label{Regge-metric}
g_{ab}^{\rm R}=\left(  
       \begin{array}{@{\,}ll}
        L_1^2 & \; F_3 \\
        F_3 & \; L_2^2 
       \end{array} 
       \\ 
 \right), \quad F_3=L_1L_2\cos \Phi_3, 
\end{equation} 
where  $\Phi_3$ is the internal angle of the vertex $1$. The triangular relation such as $L_i\!+\!L_j\!>\!L_k$ is assumed. We should note that $\sum_{i=1}^3\Phi_i$ is not always constrained to be $\pi$. For this reason, a deficit angle $\varphi$ is defined on the triangle ${\it \Delta}$ such that $\varphi=\sum_{i=1}^3 \Phi_i\!-\!\pi$ \cite{Koibuchi-2010NPB}. Equivalently, the internal angle $\Phi_i$ is obtained from  $\varphi$ such that $\Phi_i=\Phi^0_i\left(1+ {\varphi}/{\pi}\right),\,(i\!=\!1,2,3)$, where $\sum_{i=1}^3\Phi^0_i\!=\!\pi$. We assume in this paper that the variables $\Phi_i\,(i\!=\!1,2,3)$, and hence the variables $F_i\,(i\!=\!1,2,3)$, are independent from each other for the numerical simplicity. Note that $g_{ab}^{\rm R}$ reduces to the Euclidean metric $\delta_{ab}$ if $L_1\!=\!L_2\!=\!1$ and $F_3\!=\!0$.

 Note also that the Euclidean edge length $L_i$ is independent of the local coordinate, while the Regge metric $g_{ab}^{\rm R}$ itself depends on the coordinate. For this reason, the discrete Hamiltonian is obtained by using three different $g_{ab}^{\rm R}$ corresponding to three different coordinates in each triangle of the conventional model.

By replacing $L_1$ by $L_{12}$ and $L_2$ by $L_{13}$ in the Regge metric $g_{ab}^{\rm R}$ in Eq. (\ref{Regge-metric}), we get a Finsler metric $g_{ab}^{\rm F}$ such  that
\begin{equation}
\label{Finsler-metric}
g_{ab}^{\rm F}=\left(  
       \begin{array}{@{\,}ll}
        L_1^2/v_{12}^2 & \; F_3/v_{12}v_{13} \\
        F_3/v_{12}v_{13} & \; L_2^2/v_{13}^2  
       \end{array} 
       \\ 
 \right), \quad F_3=L_1L_2\cos \Phi_3. 
\end{equation} 
This $g_{ab}^{\rm F}$ gives two different lengths for each edge of the triangle. Indeed, the edge length $L_{12}(=\!L_1/v_{12})$ with respect to the local coordinate in Fig. \ref{fig-1}(a) is different from the one $L_{21}(=\!L_1/v_{21})$ with respect to the local coordinate in Fig. \ref{fig-1}(b) because $v_{12}\not=v_{21}$. This is in sharp contrast with the case of the conventional Regge metric, where the bond length of ${\it \Delta}$ is unique and independent of the coordinates. 

%++++++++++++++++++++++++++++++++++
\begin{figure}[hbt]
\centering
\includegraphics[width=8.5cm]{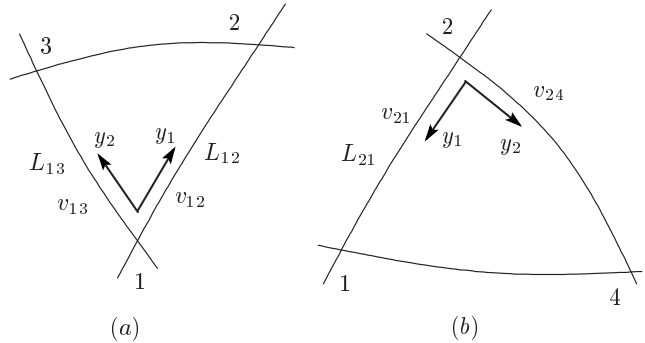}
\caption{ (a) A triangle in $M$ with the local coordinate origin is at the vertex $1$, and (b) a neighboring triangle with the local coordinate origin is at the vertex $2$. Two triangles share the bond $12$.}
\label{fig-2}
\end{figure}
%++++++++++++++++++++++++++++++++++
Figures \ref{fig-2}(a) and \ref{fig-2}(b) show two neighboring triangles which have a common bond $12$, where the vertices $1$ and $2$ are the origins of the two coordinate systems.  In this case, both $v_{12}$ and $v_{21}$ are used to define the model. The Finsler length of the bond  $12$ is given by $L_{12}$ in the triangle of Fig. \ref{fig-2}(a), while it is given by $L_{21}$ in the other triangle of Fig. \ref{fig-2}(b). This is a result of the assumption that a triangle should be labeled by a single local coordinate.

The Finsler area $A_{\it \Delta}$ of ${\it \Delta}$ is given by the determinant of $g_{ab}^{\rm F}$ such that
\begin{equation}
\label{area_of_Delta}
A_{\it \Delta}=\frac{1}{2}\sqrt{L_1^2L_2^2-F_3^2}/v_{12}v_{13}=\frac{1}{2}L_1L_2\sin \Phi_3/v_{12}v_{13}.
\end{equation} 
We should note that $A_{\it \Delta}$ depends on the local coordinate. However, this does not mean that $A_{\it \Delta}$ in Eq. (\ref{area_of_Delta}) is ill-defined. In fact, the Finsler length depends on the coordinate, so it is quite natural that $A_{\it \Delta}$ depends on the coordinate.

We should note that $g_{ab}^{\rm F}$ can also be obtained from the bi-linear form
\begin{equation}
\label{discrete-Finsler}
L_M^2=\sum_{\it \Delta}L_{\it \Delta}^2,\quad
L_{\it \Delta}\!=\!\sqrt{(L_1^2/v_{12}^2) y_1^2 \!+\!(L_1^2/v_{12}^2) y_2^2 \!+\!2(F_3/v_{12}v_{13}) y_1 y_2 }
\end{equation}
such that 
\begin{equation}
\label{g_from_L2}
g_{ab}^{\rm F}=\frac{1}{2} \frac{\partial^2 L_M^2}{\partial y_a \partial y_b}.
\end{equation}
This expression implies that $g_{ab}^{\rm F}$ is a $(0,2)$-tensor just like an ordinary metric $g_{ab}(x)$, because $y$ is a $(1,0)$-tensor and $L_M^2$ is a function.

From the $L_M$ in Eq. (\ref{discrete-Finsler}), we also have the Finsler lengths $L_{12}$ and $L_{13}$ for the bonds $12$ and $13$ of ${\it \Delta}$ in Fig. \ref{fig-1}(a). Indeed, we have $y_2=0$ ($y_1=0$) on the $x_1$ ($x_2$) axis. Thus, we find from the bi-linear form $L_{\it \Delta}^2$ that the bond length of $x_1$ ($x_2$) axis is given by $L_{12}=L_1/v_{12}$ ($L_{13}=L_2/v_{13}$). We can also start with the form $L_M^2$ in Eq. (\ref{discrete-Finsler}), because $L_{\it \Delta}^2$ can be written as $L_{\it \Delta}^2=\sum_{ab} g_{ab}^{\rm F}y_ay_b$ by using $g_{ab}^{\rm F}$ in Eq. (\ref{Finsler-metric}). 

%----------------------------------------------------------
\subsection{A surface model with Finsler metric}\label{F_metric_model}
%----------------------------------------------------------
%++++++++++++++++++++++++++++++++++
\begin{figure}[hbt]
\centering
\includegraphics[width=11cm,clip]{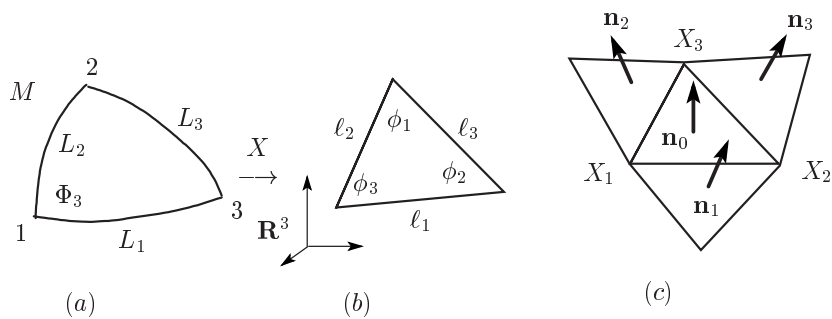}
\caption{(a) A smooth triangle ${\it \Delta}$ in $M$, (b) the image $X({\it \Delta})$ of  ${\it \Delta}$ by mapping $X$ from $M$ to ${\bf R}^3$, and (c) a unit normal vector ${\bf n}_0$ of $X({\it \Delta})$ and those ${\bf n}_i(i\!=\!1,2,3)$ of the nearest neighbor triangles in ${\bf R}^3$.}
\label{fig-3}
\end{figure}
%++++++++++++++++++++++++++++++++++
In this subsection, we define a discrete model by introducing the discrete Hamiltonian and the partition function. We assume that the surface is embedded in ${\bf R}^3$ by mapping $X: M\to {\bf R}^3$. A local coordinate is also assumed to be fixed on every triangle ${\it \Delta}$ in $M$.  

A discrete Hamiltonian $S(X,g)$ is defined by
\begin{eqnarray}
\label{Disc-Eneg} 
&&S\left(X,g\right)=S_1+\kappa S_2, \\
&&S_1=\frac {1}{4}\sum_{\it \Delta} S_1\left({\it \Delta}\right) / A_{\it \Delta},\quad
S_2=\frac {1}{ 4}\sum_{\it \Delta} S_2\left({\it \Delta}\right) / A_{\it \Delta}, \nonumber \\
&&S_1\left({\it \Delta}\right)=\left(L_1^2/v_{12}^2\right)\ell_2^2+\left(L_2^2/v_{13}^2\right)\ell_1^2-2\left(F_3/v_{12}v_{13}\right)\vec \ell_1\cdot\vec \ell_2, \nonumber \\
&&S_2\left({\it \Delta}\right)=\left(L_1^2/v_{12}^2\right)\left(1-{\bf n}_0\cdot{\bf n}_1 \right) 
                             +\left(L_2^2/v_{13}^2\right)\left(1-{\bf n}_0\cdot{\bf n}_2 \right)\nonumber \\
  &&\qquad -2\left(F_3/v_{12}v_{13}\right)\left({\bf n}_0-{\bf n}_1 \right)\cdot\left({\bf n}_0-{\bf n}_2 \right).
\nonumber
\end{eqnarray} 
The length $L_i$ and $F_3\!=\!L_1L_2\cos\Phi_3$ are the variables of  ${\it \Delta}$ in $M$ (Fig. \ref{fig-3}(a)), while the length $\ell_i$ and the unit normal vectors ${\bf n}_i(i\!=\!1,2)$ are those of $X({\it \Delta})$ in ${\bf R}^3$ (Figs. \ref{fig-3}(b),(c)). 
The coefficients $L_i^2/v_{jk}^2$ together with $1/A_{\it \Delta}$ alter the surface tension coefficient $\gamma(=\!1)$ and the bending rigidity $\kappa$ and make them dependent on the position and direction on the surface. If $v_{ij}$ is random and hence isotropic, the effective surface tension and bending rigidity are expected to be almost uniform and not anisotropic. To the contrary, an anisotropic $v_{ij}$ is expected to make these coefficients anisotropic.  

The discrete partition function is defined by
\begin{equation} 
\label{Part-Func}
 Z(\kappa) =  \int {\ D}g\int^\prime \prod _{i=1}^{N} d X_i \exp\left[-S(X,g)\right],
\end{equation} 
where
\begin{eqnarray}
\label{Disc-measure} 
\int {\ D}g=&&\sum_{\mathcal C}\int\prod _{i=1}^{N_B} d L_i\exp\left(-\lambda_{L}\sum_{i=1}^{N_B}L_i^2\right)
 \nonumber \\
&&
\int\prod _{i=1}^{N_T} d F_i\int\prod _{i=1}^{N_B} d v_i \exp\left(-\lambda_v\sum_{i}1/v_i^2\right). 
\end{eqnarray}
$\int^\prime \prod _{i=1}^{N} d X_i$ in $Z(\kappa)$ denotes that the integrations are performed under the constraint that the center of mass of the surface $X(M)$ is fixed at the origin of ${\bf R}^3$.  As mentioned in the previous subsection, we use the variable $F$ in place of $\varphi$. The integration measure $\int\prod _{i=1}^{N_T} d \varphi_i\exp\left(-\lambda_{\varphi}\sum_{i=1}^{N_T}|\varphi_i|\right)$, which is assumed in the model of \cite{Koibuchi-2010NPB}, is replaced by $\int\prod _{i=1}^{N_T} d F_i$ with a constraint $|F_3|<L_1L_2$.  Under this constraint the variable $F$ plays the role of the deficit angle $\sum_{i=1}^3 \Phi_i\!-\!\pi$ of the triangle.  
 
The role of the factor $\exp\left(-\lambda_{L}\sum_{i=1}^{N_B}L_i^2\right)$ in Eq. (\ref{Disc-measure}), where $L_i$ is the Euclidean bond length,  is to suppress the divergence of $L$. The factor $\exp\left(-\lambda_v\sum_{i}1/v_i^2\right)$ also prevents the variable $v$ from being zero. The distribution of $\{v_i\}$ becomes random and hence defines a random vector field on the surface. If the vector field ${\bf v}$ is defined otherwise externally or dynamically, this factor may be changed.

 The symbol $\sum_{\mathcal C}$ in Eq. (\ref{Disc-measure}) denotes the sum over all possible coordinates ${\mathcal C}$. A local coordinate ${\mathcal C}$ is fixed on ${\it \Delta}$. In the conventional models such as the model in \cite{Koibuchi-2010NPB}, the Hamiltonians $S_1$ and $S_2$ are defined by including the terms that are cyclic under permutations of three different coordinates of ${\it \Delta}$, such that $1\to 2$, $2\to 3$, and $3\to 1$. A permutation of three different values of ${\mathcal C}$ is not a coordinate transformation in ${\it \Delta}$.  In fact, as ${\mathcal C}$ changes from one to another, the discrete Hamiltonians $S_1$ and $S_2$ change as well. This is true not only in the conventional model but also in the Finsler geometric model. In this sense, ${\mathcal C}$ can be viewed as a variable just like the triangulation ${\mathcal T}$. However, the dynamical triangulation $\sum_{\mathcal T}$ changes the lattice structure. Therefore,  $\sum_{\mathcal T}$ cannot be included in $\sum_{\mathcal C}$. 

We should emphasize that the integrations with respect to the variables $F_i$ and $v_i$ depend on the coordinate ${\mathcal C}$. In each local coordinate of a triangle, only one of the two variables, such as $v_{12}$ and $v_{21}$, is the integration variable. Similarly in the integrations $\int^\prime\prod _{i=1}^{N_T} d F_i$, the variable $F_i$, which represents one of three different $F$, is the integration variable.

%------------------------------------------
\subsection{Continuous surface model}
\label{cont_model}
%------------------------------------------
The surface model of Helfrich and Polyakov is defined by a mapping $X$ from $M$ to ${\bf R}^3$ such that $X:M \ni  (x_1,x_2)\mapsto X(x_1,x_2)\in {\bf R}^3$ \cite{FDAVID-SMMS2004}. The symbol $(x_1,x_2)$ denotes a local coordinate system of $M$. 

The Hamiltonian of the model is given by a linear combination of the Gaussian potential $S_1$ and the extrinsic curvature energy $S_2$ such that
\begin{eqnarray}
\label{cont_S}
&&S=S_1+\kappa S_2,  \nonumber \\
&&S_1=\int \sqrt{g}d^2x g^{ab} \partial_a X^\mu \partial_b X^\mu, \quad
S_2=\frac{1}{2}\int \sqrt{g}d^2x  g^{ab} \partial_a n^\mu \partial_b n^\mu,  
\end{eqnarray} 
where $\kappa$ is the bending rigidity. Note that the unit of $\kappa$ is $[kT]$, where $k$ and $T$ are the Boltzmann constant and the temperature, respectively. The matrix $g_{ab}(a,b\!=\!1,2)$ in $S_1$ and $S_2$ is a Riemannian metric on $M$, $g$ is the determinant of  $g_{ab}$, and $g^{ab}$ is its inverse. The symbol $n^\mu$ in $S_2$ is a unit normal vector of the surface. We should note that $S_2$ is obtained from Polyakov's action for extrinsic curvature by assuming $g_{ab}\!=\!\partial_a X^\mu \partial_b X^\mu$ \cite{POLYAKOV-NPB1986}. We here assume that {$g_{ab}$} in Eq. ({\ref{cont_S}}) is arbitrary. 

The surface model described by $S$ in Eq. (\ref{cont_S}) is in statistical mechanics defined by the partition function 
\begin{equation} 
\label{cont_part_funct}
Z(b)=\int { D}g \int { D}X \exp\left[-S(X,g)\right],
\end{equation} 
where $S(X,g)$ denotes that $S$ depends on the variables $X$ and $g$. The integration symbols $\int {D}g$ and  $\int { D}X$ denote the sum over the metric $g$ and the mapping $X$. The model is characterized by the conformal invariance and the reparametrization invariance. The first means that the action $S$ remains unchanged under a transformation $g_{ab}\to g_{ab}^\prime \!=\!fg_{ab}$ for an arbitrary positive function $f(x)$. The second means that $S$ remains unchanged under any local coordinate transformation $x\to x^\prime$. The transformation $x\to x^\prime$ changes both $g$ and $X$, while the conformal transformation only changes $g$. 

We simply deform this continuous model by replacing $g_{ab}(x)$ with a Finsler metric $g_{ab}^{\rm F}(x,y)$, which is a four-variables function. In this new model the conformal invariance is apparently preserved even when the factor $f(x)$ is replaced by $f(x,y)$. In contrast, the reparametrization invariance is not always preserved, or in other words the reparametrization for $x$ is not always extended to the one for $x,y(=\!\dot{x})$. The reason is that the parameter $y$ is not a coordinate and is only allowed to transform according to a linear transformation corresponding to a coordinate transformation of $x$. However, the Finsler metric $g_{ab}^{\rm F}(x,y)$ is formally a $(0,2)$-tensor just like $g_{ab}(x)$ as mentioned just below Eq.(\ref{g_from_L2}). Therefore, the continuous actions with  Finsler metric remain scalar and hence are well-defined as action functionals.  

The discrete Hamiltonians in Eq. (\ref{Disc-Eneg}) are obtained from the continuous actions $S_1$ and $S_2$ in Eq. (\ref{cont_S}) on the triangulated surface by the replacements $\partial_1 X^\mu \!\to\! X^\mu_2\!-\!X^\mu_1$, $\partial_2 X^\mu \!\to\! X^\mu_3\!-\!X^\mu_1$, where $X^\mu_i$ denotes the position of the vertex $i$ such that $\ell_1\!=\!|X^\mu_2\!-\!X^\mu_1|$, $\ell_2\!=\!|X^\mu_3\!-\!X^\mu_1|$ (see Figs. \ref{fig-3}(a)--\ref{fig-3}(c)). The derivatives in $S_2$ in Eq. (\ref{cont_S}) can also be discretized by $\partial_1 n^\mu \!\to\! {\bf n}_0\!-\!{\bf n}_2$, $\partial_2 n^\mu \!\to\! {\bf n}_0\!-\!{\bf n}_1$, where ${\bf n}_i(i\!=\!1,2,3)$ are the unit normal vectors shown in Fig. \ref{fig-3}(c). 

%------------------------------------------
\section{Simulation results}\label{results}
%------------------------------------------
%------------------------------------------
\subsection{Euclidean model}\label{model-1}
%------------------------------------------
The model introduced in Subsection \ref{F_metric_model} is meaningful even in the case where the velocity parameter is fixed such that $v_{ij}\!=\!1$ for all $ij$ and $g_{ab}^{\rm R}\!=\!\delta_{ab}$ or $g_{ab}^{\rm R}\!=\!\partial_a X^\mu \partial_b X^\mu$. Indeed, the models in those cases are still not always identical to the corresponding conventional models because of $\sum_{\mathcal C}$ in Eq. (\ref{Disc-measure}). Therefore, in order to see the influence of $\sum_{\mathcal C}$ on the phase structure, we study not only the non-trivial Finsler geometric model (in the following subsection) but also the most simple model with $v_{ij}\!=\!1$ and $g_{ab}^{\rm R}\!=\!\delta_{ab}$ (in this subsection). 

The so-called crumpling transition between the smooth spherical phase at high bending region and the collapsed phase at low bending region has long been studied theoretically and numerically \cite{P-L-1985PRL,DavidGuitter-1988EPL,KLEINERT-PLB1986,KANTOR-KARDAR-NELSON-PRL1986,KD-PRE2002,NISHIYAMA-PRE-2004,Kownacki-Mouhanna-2009PRE}. Both of the phases separated by this transition are isotropic in the sense that the surfaces are symmetric under arbitrary three-dimensional rotations. 

To see whether this transition is not influenced by the Finsler geometric treatment, we firstly study the most simple model, which is defined by 
\begin{eqnarray}
\label{eneg-model-1} 
&& Z(\kappa) =  \sum_{\mathcal C}\int^\prime \prod _{i=1}^{N} d X_i \exp\left[-\left(S_1+\kappa S_2\right)\right],  \nonumber \\
&&S_1=\sum_{ij} \left(X_i-X_j\right)^2, \quad S_2=\sum_{ij} \left(1-{\bf n}_i\cdot{\bf n}_j\right).
\end{eqnarray}
As mentioned above, this model is identical with the conventional model except for $\sum_{\mathcal C}$ in $Z(b)$. The Hamiltonians are disctretized on a spherical lattice, which is characterized by $(N,N_B,N_T)=(10\ell^2\!+\!2,30\ell^2,20\ell^2)$, where $\ell$ is the number of partitions of an edge of the icosahedron. 

The canonical Metropolis Monte Carlo technique is used to update the variables. The update $X^\prime=X\!+\!\delta X$ is accepted with probability ${\rm Min}[1, \exp (-\delta S^\prime)]$, $\delta S^\prime=S^\prime({\rm new})\!-\!S^\prime({\rm old})$, where $\delta X$ is a random three-dimensional vector in a small sphere. The radius of this sphere is fixed to a constant to make the acceptance rate of $X$ approximately $50\%$.   

The sum over coordinates $\sum_{\mathcal C}$ in $Z(b)$ is performed as follows: On a triangle ${\it \Delta}$, the coordinate is characterized by its origin in a triangle ${\it \Delta}$, and hence ${\it \Delta}$ has only three possible coordinates. Therefore, the current coordinate of ${\it \Delta}$ is randomly updated to one of the two remaining coordinates. In this update of ${\mathcal C}$, $S_1$ and $S_2$ change from the expressions in Eq. (\ref{Disc-Eneg}) under the cyclic permutations $1\to 2$, $2\to 3$, and $3\to 1$, and so on. 

%++++++++++++++++++++++++++++++++++
\begin{figure}[hbt]
\centering
\includegraphics[width=11.0cm,clip]{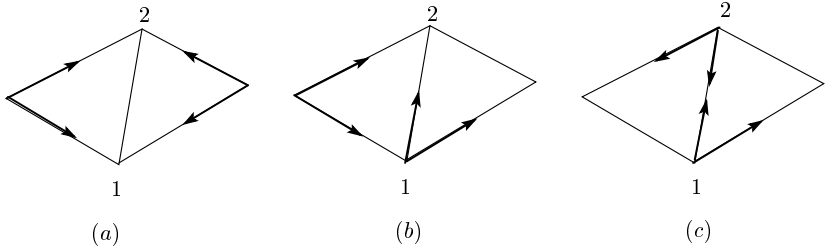}
\caption{Three configurations of coordinate axes in two neighboring triangles. Neither the bending energy nor the bond potential are assigned to the bond $12$ in the coordinate of (a), while both of the energies are redundantly assigned to the bond $12$ in the coordinate of (c). }
\label{fig-4}
\end{figure}
%++++++++++++++++++++++++++++++++++
No constraint is imposed on the update of  ${\mathcal C}$. As a consequence, the configuration like the one in Fig. \ref{fig-4}(a) appears, where the bond $12$ shares neither $S_1$ nor $S_2$. Thus, the random update of ${\mathcal C}$ makes the configuration non-uniform in the sense that the distributions of $S_1$ and $S_2$ are non-uniform. Therefore, it is not clear whether the phase structure of the conventional model is influenced by $\sum_{\mathcal C}$ in $Z$.

%++++++++++++++++++++++++++++++++++
\begin{figure}[hbt]
\centering
\includegraphics[width=13.0cm,clip]{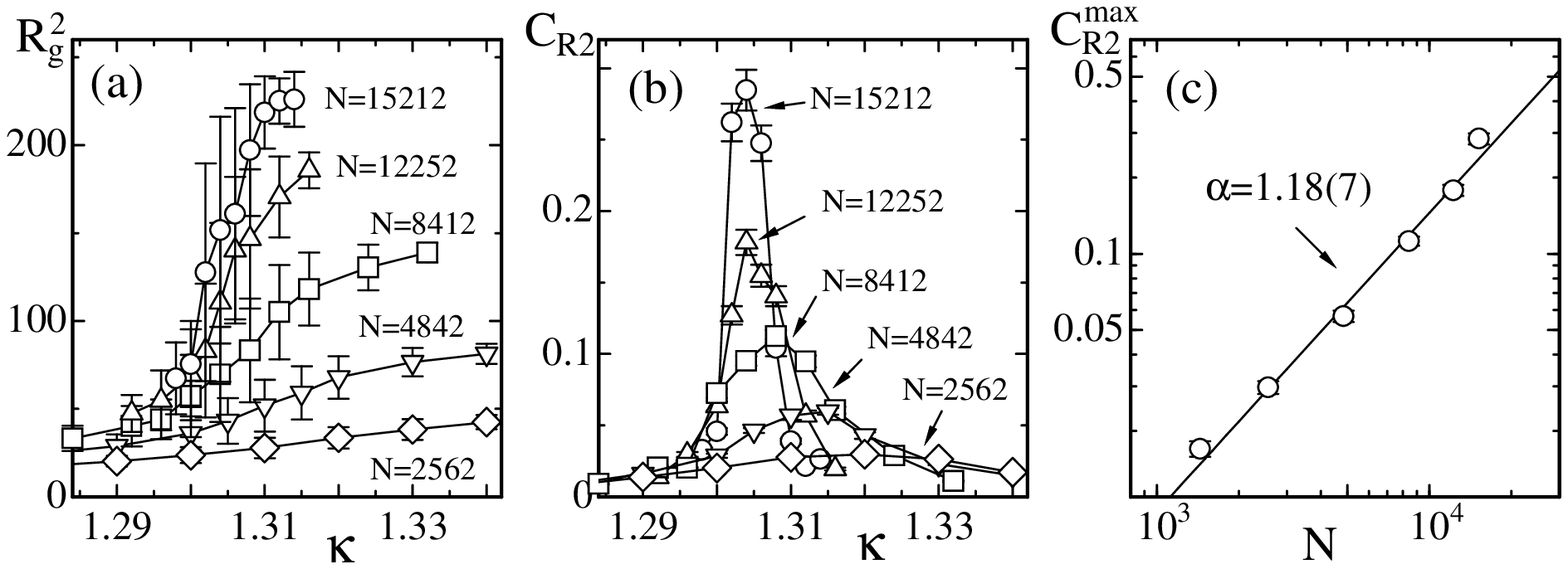}
\caption{(a) The mean square radius of gyration $R_g^2$ vs. $\kappa$, (b) the variance $C_{R^2}$ vs. $\kappa$, and (c) a log-log plot of the peak $C_{R^2}^{\rm max}$ vs. $N$. The solid lines connecting the data in (a),(b) are drawn to guide the eyes. The straight line in (c) is drawn by fitting the data to Eq. (\ref{CX2-scaling}).  }
\label{fig-5}
\end{figure}
%++++++++++++++++++++++++++++++++++
The mean square radius of gyration $R_g^2$ is defined by 
\begin{equation}
\label{X2}
R_g^2={1\over N} \sum_i \left(X_i-\bar X\right)^2, \quad \bar X={1\over N} \sum_i X_i,
\end{equation}
where $\bar X$ is the center of mass of the surface. The large errors in $R_g^2$ reflect large fluctuation of $R_g^2$ as a result of the crumpling transition between the smooth phase and the crumpled phase (Fig. \ref{fig-5}(a)). The variance $C_{R^2}$ of $R_g^2$ defined by
\begin{equation}
\label{CX2}
C_{R^2}={\frac 1 N} \left< \left(R_g^2-\langle R_g^2\rangle \right)^2 \right>
\end{equation}
can reflect the phase transition. The peak at $\kappa\!\simeq\!1.303$ in Fig. \ref{fig-5}(b) indicates the existence of the transition. The peak values $C_{R^2}^{\rm max}$ scales against $N$; the straight line in Fig. \ref{fig-5}(c) is drawn by fitting the data to 
\begin{equation}
\label{CX2-scaling}
C_{R^2}^{\rm max}\sim N^{\alpha}, \quad \alpha=1.18\pm 0.07.
\end{equation}
The obtained exponent $\alpha\!=\!1.18(7)$ indicates that the transition is of first-order.  

%++++++++++++++++++++++++++++++++++
\begin{figure}[hbt]
\centering
\includegraphics[width=13.0cm,clip]{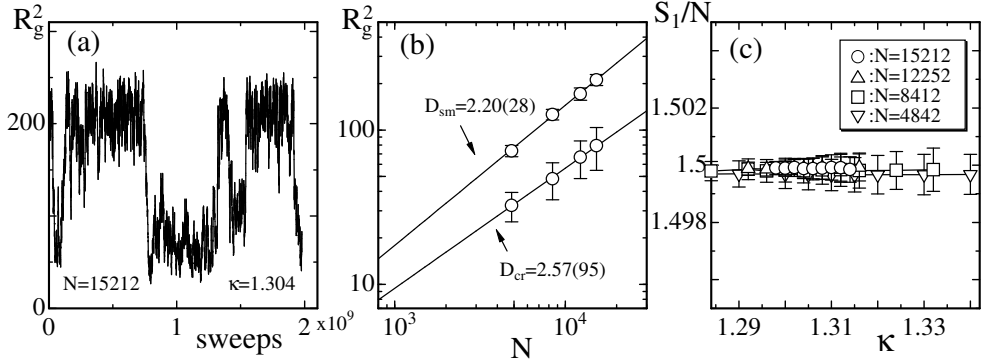}
\caption{(a) The variation of $R_g^2$ against MCS at the transition on the $N\!=\!15212$ surface, (b) log-log plots of $R_g^2$ vs. $N$ in the smooth and crumpled phases at the transition point, (c) $S_1/N$ vs. $\kappa$.   }
\label{fig-6}
\end{figure}
%++++++++++++++++++++++++++++++++++
 We find from the plot of the series $\{R_g^2\}$ in Fig. \ref{fig-6}(a) that the smooth and crumpled phases are clearly separated. This series is obtained at the transition point $\kappa\!=\!1.304$ on the $N\!=\!15212$ surface. Plots of $\{R_g^2\}$ similar to the one in Fig. \ref{fig-6}(a) are obtained on the $N\!=\!12252$ and $N\!=\!8412$ surfaces, though they are not depicted.  In order to have the fractal dimension $D_{\rm f}$ defined by $R_g^2\sim N^{2/D_{\rm f}}$, we calculate the mean values of $R_g^2$ in the smooth and crumpled phases independently from the series  $\{R_g^2\}$. On the surfaces $N\!\geq\!8412$, we use the series $\{R_g^2\}$ at the transition point like the one in Fig. \ref{fig-6}(a), while on the surfaces $N\!\leq\!4842$ we use two different $\{R_g^2\}$ obtained in the smooth and crumpled phases. Figure \ref{fig-6}(b) shows the results $R_g^2$ vs. $N$ in the log-log scale. From the slope of the fitted lines, we have
\begin{equation}
\label{H-Eucl}
D_{\rm sm}=2.20\pm 0.28\; ({\rm smooth}),\quad D_{\rm cr}=2.57\pm 0.95\; ({\rm crumpled}).
\end{equation}
These values are comparable with the results $D_{\rm sm}\!=\!2.02(14)$ and $D_{\rm cr}\!=\!2.59(57)$ of the conventional model within the errors \cite{Koibuchi-2005PRE}. 

We see the expected relation $S_1/N\!=\!1.5$ in Fig. \ref{fig-6}(c). This implies that the equilibrium configurations are correctly obtained under  $\sum_{\mathcal C}$ in $Z$.

%++++++++++++++++++++++++++++++++++
\begin{figure}[hbt]
\centering
\includegraphics[width=13.0cm,clip]{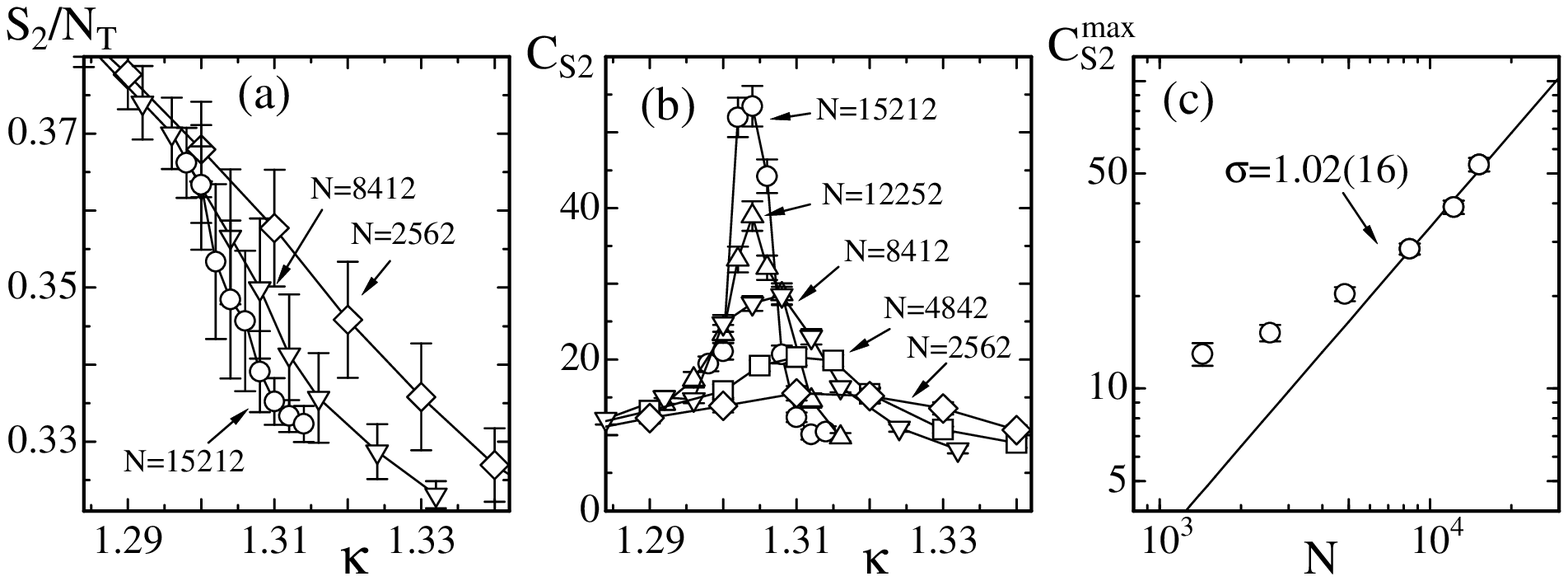}
\caption{(a) The bending energy $S_2/N_T$ vs. $\kappa$, (b) the specific heat $C_{S_2}$ vs. $\kappa$, and (c) the log-log plot of the peak values $C_{S_2}^{\rm max}$ vs. $N$. The fitting is performed by using the largest three data in (c).  }
\label{fig-7}
\end{figure}
%++++++++++++++++++++++++++++++++++
The bending energy $S_2/N_T$, the specific heat 
\begin{equation}
\label{CS2}
C_{S_2}=\frac {\kappa^2} {N} \left< \left(S_2-\langle S_2\rangle \right)^2 \right>,
\end{equation}
and the peak values $C_{S_2}^{\rm max}$ are plotted in Figs. \ref{fig-7}(a)--\ref{fig-7}(c). By fitting the largest three values of $C_{S_2}^{\rm max}$, we have 
\begin{equation}
\label{CS2-scaling}
C_{S_2}^{\rm max}\sim N^{\sigma}, \quad \sigma=1.02\pm 0.16.
\end{equation}
The exponent $\sigma\!=\!1.02(16)$ also confirms that transition is of first-order, and the value is in good agreement with $\sigma\!=\!0.93(13)$ of the conventional model \cite{Koibuchi-2005PRE}.

%------------------------------------------
\subsection{Regge metric model: random vector field}\label{model-2}
%------------------------------------------
In this subsection, we study a non-trivial model, which is defined by 
\begin{eqnarray}
\label{eneg-model-2} 
&&Z(\kappa) =  \int { D}g\int\prod_{i=1}^{N_T} d\rho_i\int^\prime \prod _{i=1}^{N} d X_i \exp\left[-\left(S_1+\kappa S_2+\lambda_3 S_3+\lambda_3 S_4\right)\right],  \nonumber \\
&&\int {\ D}g=\sum_{\mathcal C}\int\prod _{i=1}^{N_B} d L_i\exp\left(-\lambda_{L}\sum_{i=1}^{N_B}L_i^2\right) \int\prod _{i=1}^{N_B} d v_i \exp\left(-\lambda_v\sum_{i}1/v_i^2\right), \nonumber \\ 
&&S_1 \;{\rm and}\; S_2 ={\rm Eq.}(\ref{Disc-Eneg}), \\
&&S_3= \sum_{i=1}^{N_T} \rho_iA_i, \quad S_4=\sum_{ij}|\rho_i-\rho_j|.\nonumber
\end{eqnarray}
In this model, the variable $\varphi$ for the deficit angle is omitted for simplicity, and hence the variable $F$ in Eq. (\ref{Disc-measure}) is only given by $F_3\!=\!\left(L_1^2\!+\!L_2^2\!-\!L_3^2\right)/2$. In this case the metric $g_{ab}^R$ in  Eq.(\ref{Regge-metric}) is identical with the conventional Regge metric \cite{REGGE-NC1961,HAMBER-LH1986,FDAVID-LH1992}.

The symbol $\rho_{i}$ in $S_3$ denotes the scalar field on $M$, which is the conjugate variable to the surface area $A_{i}$, and $S_4$ is the interaction term. $\sum_{ij}$ in $S_4$ is the sum over all nearest neighbor triangles $i$ and $j$. The coefficients $\lambda_3$ and $\lambda_4$ are fixed to $\lambda_3\!=\!\lambda_4\!=\!1$. The variable $\rho_{i}$ and the terms $S_3$ and $S_4$ are not always necessary but they are introduced to take the in-plane deformation into account. In this model, $\rho$ interacts with the surface through the coupling $\rho_iA_i$. To the contrary,  a constant scalar field $\rho$ has no explicit interaction with the surface. If $A_i$ is constant, then the field $\rho$ and hence both $S_3$ and $S_4$ are independent of the surface geometry. 

The effective Hamiltonian $S^\prime$ including the measure terms is given by  $S^\prime=S_1\!+\!\kappa S_2\!+\!\sum_i L_i^2 \!+\! \sum_i {\tilde v}_i^2\!+\!S_3\!+\!S_4$, where $L_i$ is the Euclidean bond length.
In this expression, we replace $1/v_i$ with the inverse velocity  ${\tilde v}_i=1/v_i$ for numerical simplicity. The variables summed over in the partition function are $X$, $L$, ${\tilde v}$, and ${\mathcal C}$. One MCS consists of $N$ updates of $X$, $N_B$ updates of $L$, $N_B$ updates of ${\tilde v}_i$, and $N_T$ updates of ${\mathcal C}$.     

The variable $L$ is updated in such a way: $L^\prime=L\!+\!\delta L (>0)$ with random numbers $\delta L\in [-0.5,0.5]$. In this update, $L^\prime$ is constrained to satisfy the triangle equalities. The inverse velocity ${\tilde v}_i(i=1,2)$ is updated such that ${\tilde v}_i^\prime={\tilde v}_i\!+\!\delta {\tilde v}_i(>\!0)$ with a random number $\delta {\tilde v}_i\in [-0.5,0.5]$. None of the variables ${\tilde v}_i$ is updated on the bond $12$ of the configuration in Fig. \ref{fig-4}(a), one of ${\tilde v}_i$ is updated in Fig. \ref{fig-4}(b), and both of the variables are updated in Fig. \ref{fig-4}(c). The constraint ${\tilde v}_i<1$ is imposed. Without this constraint, the acceptance rate for the update of local coordinate ${\mathcal C}$ remains very small ($10\%\sim 15\%$), while it remains $40\%\sim 50\%$ under the constraint. 

%++++++++++++++++++++++++++++++++++
\begin{figure}[hbt]
\centering
\includegraphics[width=13.0cm,clip]{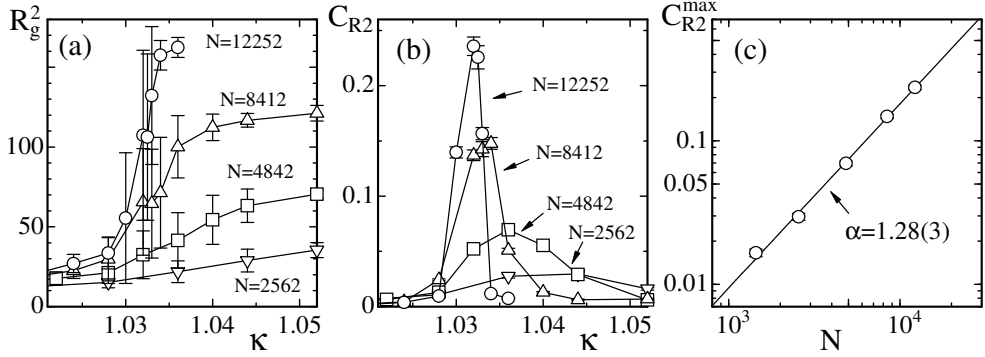}
\caption{(a) The mean square radius of gyration $R_g^2$ vs. $\kappa$, (b) the variance $C_{R^2}$ vs. $\kappa$, and (c) the log-log plot of the peak values $C_{R^2}^{\rm max}$ vs. $N$, where $\lambda_3\!=\!\lambda_4\!=\!1$.  }
\label{fig-8}
\end{figure}
%++++++++++++++++++++++++++++++++++
The mean square radius of gyration $R_g^2$ against $\kappa$ in Fig. \ref{fig-8}(a) is almost identical to that of the Euclidean model in the previous subsection. The variance $C_{R^2}$ and the peak value  $C_{R^2}^{\rm max}$ shown in Figs. \ref{fig-8}(b) and \ref{fig-8}(c) are also almost identical to those of the previous subsection. The straight line in Fig. \ref{fig-8}(c) is drawn by fitting the data to Eq. (\ref{CX2-scaling}), and we have $\alpha=1.28\pm0.03$. This value is identical to the one in Eq. (\ref{CX2-scaling}) within the error. 

%++++++++++++++++++++++++++++++++++
\begin{figure}[hbt]
\centering
\includegraphics[width=13.0cm]{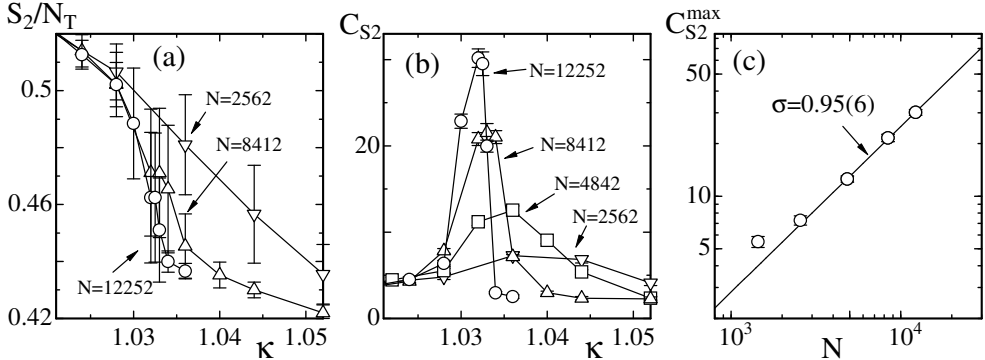}
\caption{(a) The bending energy $S_2/N_T$ vs. $\kappa$, (b) the specific heat $C_{S_2}$ vs. $\kappa$, and (c) the log-log plot of the peak values $C_{S_2}^{\rm max}$ vs. $N$, where $\lambda_3\!=\!\lambda_4\!=\!1$.  }
\label{fig-9}
\end{figure}
%++++++++++++++++++++++++++++++++++
The bending energy $S_2/N_B$ and the specific heat $C_{S_2}$ are almost identical with those of the Euclidean model (Figs. \ref{fig-9}(a),(b)). The scaling of $C_{S_2}^{\rm max}$ predicted in Eq. (\ref{CS2-scaling}) gives the exponent  $\sigma\!=\!0.95(6)$, which is almost comparable to that of the Euclidean model. The fractal dimension $D_{\rm f}$ is calculated from the series of $\{R_g^2\}$ at the transition, and the results are $D_{\rm sm}\!=\!2.12(20)$ (smooth) and $D_{\rm cr}\!=\!3.0(10)$ (crumpled). The result $D_{\rm sm}\!=\!2.12(20)$ is comparable with the one in the previous subsection, while $D_{\rm cr}\!=\!3.0(10)$ is slightly larger than the corresponding result in the previous subsection. However, we see no difference in the phase structures between the Finsler and conventional models. This implies that the Finsler geometric treatments including $\sum_{\mathcal C}$ are well-defined.

We performed the simulations for a model with the variable $F$, which corresponds to the deficit angle $\varphi$ of the triangles in $M$. This model is identical with the one in this subsection except the variable $F$. The results are consistent with those of the conventional model just like the models shown in this and the preceding subsections. 

%------------------------------------------
\subsection{Anisotropic surface model: constant vector field}\label{model-3}
%------------------------------------------
In this subsection, we see that a tubular surface is obtained under a constant vector field on the surface. Let $g_{ab}^R$ in Eq. (\ref{Regge-metric}) be $g_{ab}^R\!=\!\delta_{ab}$ and  $|v_x|\!=\!v_{12}\!=\!v$, $|v_y|\!=\!v_{13}\!=\!1$, where $(x,y)$ is a local coordinate along ${\bf v}$ on $M$, then we have {$S_2\!=\!(1/2)\int dxdy \left[(1/v)(\partial_x n^\mu)^2\!+\!v(\partial_y n^\mu)^2\right]$}. Thus we have an anisotropic bending rigidity such that $\kappa_x\!=\!\kappa/v$ and $\kappa_y\!=\!\kappa v$. 
If $v\!<\!1$, then we have $\kappa_x\!>\!\kappa_y$, and consequently the surface becomes smooth (wrinkled) in the $x$ direction ($y$ direction) in a certain range of $\kappa$. In this model, a tubular surface is expected at sufficiently large or small $v$, although none of the parameters $\kappa_x$, $\kappa_y$ can be exactly 0.

Thus we have anisotropic bending rigidities $\kappa_x$ and $\kappa_y$ if the vector field ${\bf v}$ is constant. Since a vector field ${\bf v}$ on $M$ corresponds to the one on the surface $X(M)\subset{\bf R}^3$, we assume a constant in-plane tilt order ${\vec \sigma}$ at the center of each triangle. This in-plane variable ${\vec \sigma}$ becomes the vector field on $X(M)$ and corresponds to ${\bf v}$ in $M$. 

The variable ${\vec \sigma}$ is of unit length $|{\vec \sigma}|\!=\!1$ and has a value in ${\bf R}^3$. 
%++++++++++++++++++++++++++++++++++
\begin{figure}[hbt]
\centering
\includegraphics[width=9.5cm,clip]{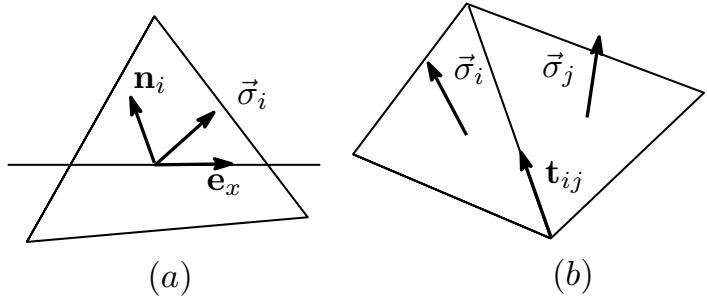}
\caption{(a) The $x$-direction component $\vec \sigma_i$ defined by Eq. (\ref{sigma}), the unit normal vector ${\bf n}_i$, and the canonical basis vector ${\bf e}_x\!=\!(1,0,0)$, of the triangle $i$, (b) the variables $\vec \sigma_i$ and $\vec \sigma_j$ of the triangles $i$ and $j$, and the unit tangential vector ${\bf t}_{ij}$ of the bond $ij$.  }
\label{fig-10}
\end{figure}
%++++++++++++++++++++++++++++++++++
The direction of ${\vec \sigma}_i$ is defined by the projection of ${\bf e}_x\!=\!(1,0,0)\in{\bf R}^3$ on the triangle plane (Fig. \ref{fig-10}(a)) such that
\begin{equation}
\label{sigma} 
\vec \sigma_i=\frac{{\bf e}_x-({\bf e}_x\cdot{\bf n}_i){\bf n}_i}{|{\bf e}_x-({\bf e}_x\cdot{\bf n}_i){\bf n}_i|},
\end{equation}
where ${\bf n}_i$ is the unit normal vector. As the surface shape varies, not only ${\bf n}_i$ but also ${\vec \sigma}_i$ varies. We use the word "constant" in the sense that  ${\vec \sigma}_i$ is defined by the constant vector ${\bf e}_x\!=\!(1,0,0)$. This ${\vec \sigma}_i$ represents the constant in-plane tilt order.

The variable ${\vec \sigma}_i$ plays a role of ${\bf v}$ on the surface, and therefore the component ${\vec \sigma}_i\cdot {\bf t}_{ij}$ is used to define the bending rigidity at the bond $ij$, where ${\bf t}_{ij}$ is a unit tangential vector along the bond $ij$. The effective bending rigidity $\kappa_{ij}$ is given by {$\kappa_{12}\!=\!\kappa v_{13}v_{12}^{-1}$} and {$\kappa_{13}\!=\!\kappa v_{12}v_{13}^{-1}$}, where {$v_{12}$} and {$v_{13}$} are the vectors along the bonds {$12$} and {$13$} in Fig. {\ref{fig-1}(a)}. However, this {$\kappa_{ij}$} becomes singular if $v_{ij}\!=\!0$. For this reason, we simply define {$v_{ij}$} by multiplying the integer {$N_v$} so that {$v_{ij}$} has an integer value in {$\{1,2,\cdots, N_v\}$}: 
\begin{equation}
\label{anisotropic-v} 
v_{ij}= 1+\left[\sigma_{ij}\right],\quad \sigma_{ij}=N_v \left|{\vec \sigma}_i\cdot {\bf t}_{ij}\right|, \qquad ({\rm case\; 1}),
\end{equation} 
where {$[x]$} represents the integer {${\rm Max}\{n\in {\bf Z}|n\leq x\}$}. Consequently, the minimum (maximum) value of the effective bending rigidity becomes {$\kappa \;(N_v \kappa)$}. 

Another possible discretization of {$\kappa_{ij}$} is given by modifying  $\kappa_{ij}$ from the naive discretization $\kappa_{12}\!=\!\kappa v_{13}v_{12}^{-1}$ and $\kappa_{13}\!=\!\kappa v_{12}v_{13}^{-1}$ to $\kappa_{12}\!=\!\kappa v_{12}^{-1}$ and $\kappa_{13}\!=\!\kappa v_{13}^{-1}$,  and by defining $v_{ij}^{-1}$ such that
\begin{equation}
\label{anisotropic-b} 
v_{ij}^{-1}= \left\{
       \begin{array}{@{\,}ll}
        1 & \quad \left( |{\vec \sigma}_i\cdot {\bf t}_{ij}|<\cos \theta_0\right) \\
        c_0 & \quad \left( |{\vec \sigma}_i\cdot {\bf t}_{ij}|\geq \cos \theta_0\right) 
       \end{array} 
       \right.,\quad (\theta_0=\frac{\pi}{3})\qquad ({\rm case\; 2}),\\ 
\end{equation} 
where $c_0(<\!1)$ is a constant to be fixed and $\theta_0$ is also fixed to yield $\cos \theta_0\!=\!0.5$. It is possible to define {$v_{ij}^{-1}$} in Eq.({\ref{anisotropic-b}}) by using $(1/2)({\vec \sigma}_i+{\vec \sigma}_j)\cdot {\bf t}_{ij}$. In this case, {$v_{ij}^{-1}$} is uniquely defined on the bond $ij$ ($v_{ij}\!=\!v_{ji}$), and hence it is possible to assume the conventional discretization for $S_2$, where $\sum_{\mathcal C}$ is not included in the partition function $Z$. However, we use {$v_{ij}^{-1}$} in Eq. (\ref{anisotropic-b}) as a demonstration, where $\sum_{\mathcal C}$ is included in $Z$. 

The partition function $Z$, the Gaussian bond potential $S_1$ and the bending energy $S_2$ are given by
\begin{eqnarray}
\label{eneg-model-3} 
&& Z(b) =  \sum_{\mathcal C}\int^\prime \prod _{i=1}^{N} d X_i \exp\left[-\left(S_1+\kappa S_2\right)\right], \nonumber \\
&&S_1=\sum_{\it \Delta} S_1({\it \Delta}),\quad S_2=\sum_{\it \Delta} S_2({\it \Delta}), \\ 
&&S_1({\it \Delta})=\left(X_2-X_1\right)^2+\left(X_3-X_1\right)^2,  \nonumber \\
&&S_2({\it \Delta})=\frac{v_{13}}{v_{12}}\left(1-{\bf n}_0\cdot{\bf n}_1\right)+\frac{v_{12}}{v_{13}}\left(1-{\bf n}_0\cdot{\bf n}_2\right), \qquad ({\rm case\; 1}), \nonumber  \\
&&S_2({\it \Delta})=v_{12}^{-1}\left(1-{\bf n}_0\cdot{\bf n}_1\right)+v_{13}^{-1}\left(1-{\bf n}_0\cdot{\bf n}_2\right), \qquad ({\rm case\; 2}), \nonumber
\end{eqnarray}
where the surface tension coefficients {$\gamma_{12}\!=\!v_{12}v_{13}^{-1}$} and {$\gamma_{13}\!=\!v_{13}v_{12}^{-1}$} of $S_1({\it \Delta})$ are  fixed to $1$ for simplicity. The unit normal vectors {${\bf n}_0,{\bf n}_1,{\bf n}_2$} in {$S_2({\it \Delta})$} are those shown in Fig. {\ref{fig-3}}(c). Due to the coefficient $\kappa_{ij}${($=v_{13}v_{12}^{-1},\cdots,v_{13}^{-1}$)} in $S_2({\it \Delta})$, we have the effective bending rigidity $\kappa\kappa_{ij}$ as mentioned above.  

%++++++++++++++++++++++++++++++++++
\begin{figure}[hbt]
\centering
\includegraphics[width=10.0cm,clip]{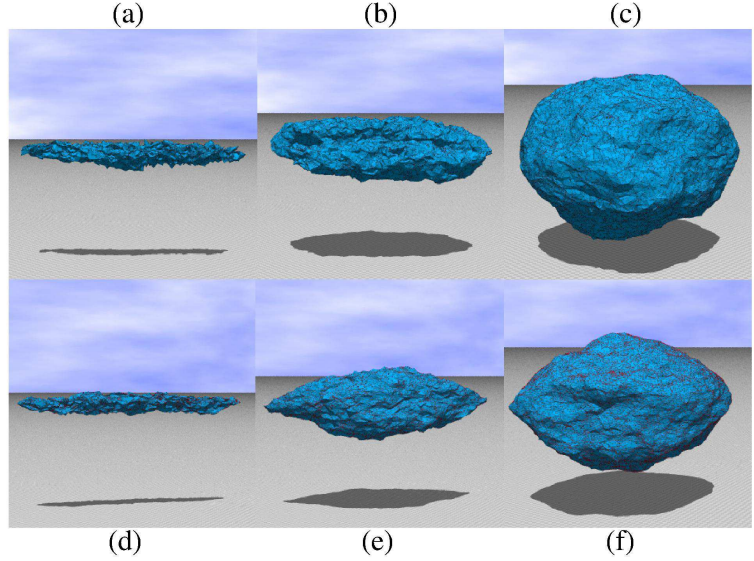}
\caption{(Color online) Snapshot of surfaces of size $N\!=\!10242$ of case 1 model obtained for (a) {$\kappa\!=\!1$}, (b) {$\kappa\!=\!2$}, and (c) {$\kappa\!=\!5$}, where {$N_v\!=\!10$}. Snapshots of case 2 model obtained for (d) {$(c_0,\kappa)\!=\!(0.2,3.5)$}, (e) {$(c_0,\kappa)\!=\!(0.2,7.5)$} and (f) {$(c_0,\kappa)\!=\!(0.2,15)$}. }
\label{fig-11}
\end{figure}
%++++++++++++++++++++++++++++++++++
The integer {$N_v$} in Eq.(\ref{anisotropic-v}) is fixed to {$N_v\!=\!10$} in case 1, and the parameter $c_0$ in Eq.(\ref{anisotropic-b}) is fixed to  $c_0\!=\!0.2$ in case 2, while $\kappa$ is varied in the simulations. Snapshots in Figs.\ref{fig-11}(a)--(f) show that tubular surfaces are obtained in both case 1 ((a)--(c)) and case 2 ((d)--(f)).  Since a vector field on a sphere has singular points, the variable $\sigma_i$ becomes singular on the surface. Under the condition of Eq. (\ref{sigma}), we expect that there appear two singular points. Indeed, these two points can easily be seen at two terminal points of the surface in the snapshots at relatively small {$\kappa$} in both case 1 and case 2.   

%------------------------------------------
\subsection{Anisotropic surface model: dynamical vector field with the Heisenberg spin model Hamiltonian}\label{model-4}
%------------------------------------------
In this subsection, the vector field ${\bf v}$ is assumed to be dynamically changed. The variable ${\vec \sigma}$ is defined at the vertices of triangles as the tilt variable such that its in-plane components play the role of the vector filed ${\bf v}$ on the surface \cite{Koibuchi-Msquare-2012}. Interaction between the tilts ${\vec \sigma}$ is included in the Hamiltonian. This interaction is a polar one and hence it does not always represent the interaction of the liquid crystal molecules in LCEs membranes. The fictitious variable ${\mathcal C}$ is simply summed over in the discrete Hamiltonian, and hence it is not included in the sum of partition function just like in the conventional treatment of the surface models. 

The Hamiltonian is given by 
\begin{eqnarray}
\label{Disc-Eneg-4} 
S(X,\sigma)= \lambda S_0+ S_1 + \kappa S_2, 
\end{eqnarray} 
where $S_0$ is the energy for the tilts $\sigma_i (\in {\bf S}^2$:unit sphere) and is given by $
S_0\!=\!({1}/{2}) \int \sqrt{g}g^{ab}({\partial \sigma}/{\partial x_a})\cdot({\partial \sigma}/{\partial x_b}).
$
The metric function in $S_1$ and $S_2$ is replaced by $g_{ab}^F$ in Eq.(\ref{Finsler-metric}) with $L_1\!=\!L_2\!=\!1$ and $F_3\!=\!0$, while the metric in $S_0$ is fixed to be $\delta_{ab}$.

The partition function $Z$, and the discrete energies  $S_0$, $S_1$, and $S_2$ are defined as follows: 
\begin{eqnarray}
\label{Disc-Eneg-5} 
&& Z(\lambda,\kappa) =  \sum_\sigma\int^\prime \prod _{i=1}^{N} d X_i \exp\left[-\left(\lambda S_0+S_1+\kappa S_2\right)\right], \nonumber \\
&&S_0=\sum_{ij}\left( 1-\sigma_i\cdot \sigma_j\right), \quad
S_1=\sum_{\Delta}S_{1\Delta}, \quad S_2=\sum_{\Delta}S_{2\Delta}, \nonumber \\
&&S_{1\Delta}=  \frac{1}{6} \left(\gamma_{1}\ell_1^2+\gamma_{2}\ell_2^2+\gamma_{3}\ell_3^2\right), \\
&&S_{2\Delta}=\frac{1}{6} \left[ \kappa_{1}(1-{\bf n}_0\cdot{\bf n}_1)  +\kappa_{2}(1-{\bf n}_0\cdot{\bf n}_2)+\kappa_{3}(1-{\bf n}_0\cdot{\bf n}_3)\right],\nonumber
\end{eqnarray} 
where $\ell_i(i=1,2,3)$ is the bond length of the triangles $\Delta$,  and ${\bf n}_i(i=0,1,2,3)$ denote the unit normal vectors of triangles (Fig. \ref{fig-3}(c)). $\sum_{\Delta}$ denotes the sum over all $\Delta$. 

The coefficients $\gamma_{i}$ and $\kappa_{i}$ in Eq. (\ref{Disc-Eneg-5}) are defined by
\begin{eqnarray}
\label{coefficients-2}
&&\gamma_{1}=v_{13}^{-1}+v_{23}^{-1},\quad \gamma_{2}=v_{12}^{-1}+v_{32}^{-1},\quad \gamma_{3}=v_{21}^{-1}+v_{31}^{-1}, \nonumber \\
&&\kappa_{1}=v_{12}^{-1}+v_{21}^{-1},\quad \kappa_{2}=v_{13}^{-1}+v_{31}^{-1},\quad \kappa_{3}=v_{23}^{-1}+v_{32}^{-1}, \nonumber \\
&&\quad v_{ij}^{-1}= \left\{\begin{array}{@{\,}ll}
   1 & \quad (|\sigma_i\cdot{\bf t}_{ij}|<\cos\theta_0) \\
   c_0 \;(<1) & \quad (|\sigma_i\cdot{\bf t}_{ij}|\geq\cos\theta_0) 
   \end{array} 
   \right.,\quad (\theta_0=\frac{\pi}{4}). 
\end{eqnarray}
This definition implies that $\gamma_{i}$ and $\kappa_{i}$ have values in $\{2,2c_0,1\!+\!c_0\}$. If all $\sigma_i$ satisfy  $|\sigma_i\cdot{\bf t}_{ij}|<\cos\theta_0$, then we have $\gamma_{i}\!=\!\kappa_{i}\!=\!2$ and therefore $S_1$ and $S_2$ reduces to the ordinary ones such as $S_1\!=\!\sum_{ij}\ell_{ij}^2$, $S_2\!=\!\sum_{ij}\left(1\!-\!{\bf n}_i\cdot{\bf n}_j\right)$ up to the multiplicative factor $2/3$. We should note that $\gamma_i$ becomes the effective surface tension while $\kappa\kappa_i$ becomes the effective bending rigidity at the bond $i$. Thus, the surface tension and bending rigidity dynamically changes depending on the position and the direction on the surface.

%++++++++++++++++++++++++++++++++++
\begin{figure}[!h]
\centering
\includegraphics[width=12cm,clip]{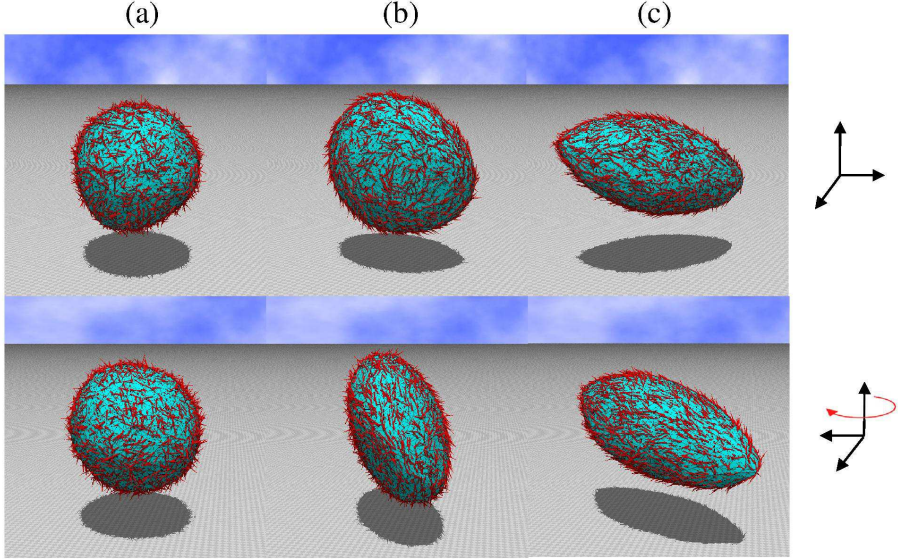}  
\caption{(Color online) Snapshots obtained for (a) $\lambda\!=\!0.56$ (spherical),  (b) $\lambda\!=\!0.68$ (disk), (c) $\lambda\!=\!1$ (tubular). $N\!=\!2562$, $\kappa\!=\!320$, and $(c_0,\theta_0)\!=\!(0.2,\pi/4)$ on connection-fixed surfaces. The  brushes on the surface represent the tilt variables. The view point of the lower snapshot is rotated $\pi/2$ around the vertical axis of the upper one in (a)--(c). 
} 
\label{fig-12}
\end{figure}
%++++++++++++++++++++++++++++++++++
Snapshots of connection-fixed surfaces of size $N\!=\!2562$ are shown in Figs. \ref{fig-12}(a)--(c). We see that there are three different phases; spherical phase, disk phase and tubular phase.  Brushes on the surface denote the tilts. The  directions of tilts are at random in the spherical phase (Fig. \ref{fig-12}(a)),  and aligned in the tubular phase (Fig. \ref{fig-12}(c)). In the disk phase in Fig. \ref{fig-12}(b), the tilts form a vortex-like configuration just like the Kosterlitz-Thouless phase in the two-dimensional $XY$ model, where $\sigma_i (\in {\bf S}^1$:unit circle).

%++++++++++++++++++++++++++++++++++
\begin{figure}[!h]
\centering
\includegraphics[width=13.5cm]{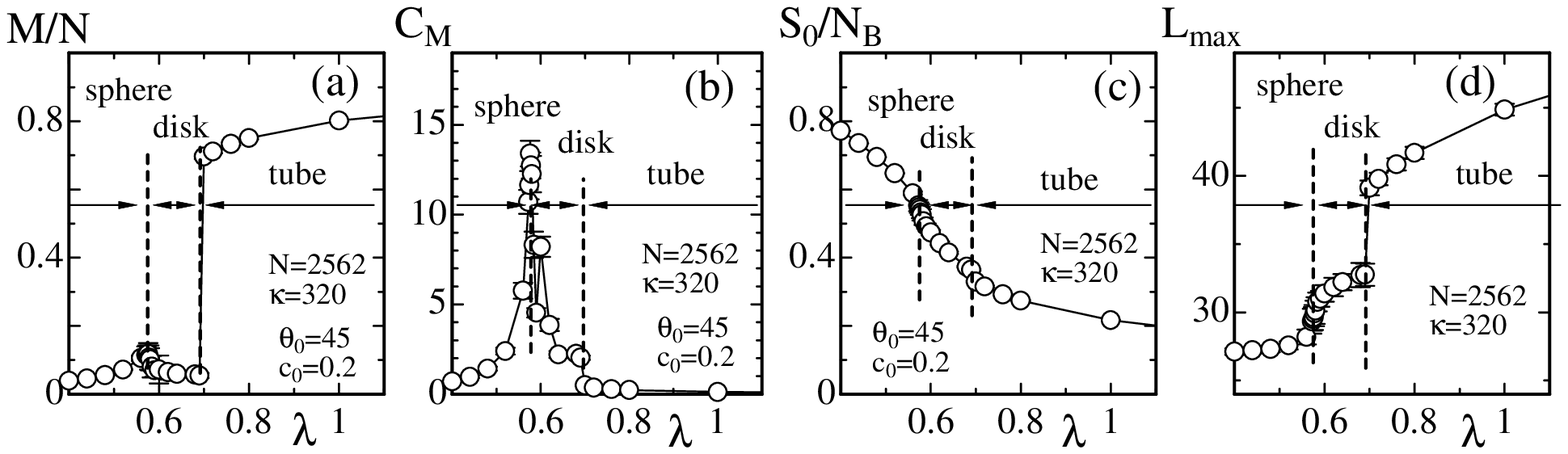}  
\caption{ (a) $M/N$ vs. $\lambda$, (b) $C_M$ vs. $\lambda$, (c) $S_0/N_B$ vs. $\lambda$, and (d) the maximal linear extension $L_{\rm max}$ vs. $\lambda$. $N\!=\!2562$, $\kappa\!=\!320$, and $(c_0,\theta_0)\!=\!(0.2,\pi/4)$.  } 
\label{fig-13}
\end{figure}
%++++++++++++++++++++++++++++++++++
The magnetization $M$ defined by 
\begin{eqnarray} 
\label{M}
M=\frac{1}{N}\left|\sum_i\sigma_i\right|=\frac{1}{N}\left|\left(\sum_i \sigma_i^x,\sum_i \sigma_i^y,\sum_i \sigma_i^z\right)\right|  
\end{eqnarray} 
discontinuously changes against $\lambda$ at the phase boundary between the disk and tubular phases as shown in Fig.\ref{fig-13}(a). This indicates that the tilt configuration is reflected in the surface shape. The variance $C_M\!=\!(1/N)\langle \left(M\!-\!\langle M\rangle\right)^2 \rangle$ has a peak at the boundary between the disk and spherical phases (Fig. \ref{fig-13}(b)). The  energy $S_0/N_B$ rapidly changes at the same boundary. This implies a phase transition (Fig. \ref{fig-13}(c)), however detailed analyses are not yet performed. The maximal linear extension $L_{\rm max}$ of the surface changes discontinuously (almost discontinuously) at the boundary between the disk and tubular (spherical) phases (Fig. \ref{fig-13}(d)). Thus, we confirm that an internal in-plane or external tilt order is a possible origin of anisotropy, although the lattice size used here is not so large.

It is also possible to define $v_{ij}$ such that  $v_{ij}\!=\!\left|\sigma_i\cdot{\bf t}_{ij}\right|\!+\!\epsilon$, where small number $\epsilon$ is introduced to prevent $v_{ij}^{-1}$ from being divergent. The Finsler metric can also be introduced in $S_0$.  
These problems remain to be studied in future. 

%------------------------------------------
\section{Summary and Conclusion}\label{Conclusion}
%------------------------------------------
In this paper, we have constructed a Finsler geometric (FG) surface model on the triangulated surfaces by including the three dimensional tilt variable ${\vec \sigma}$ into the metric function, which is a Finsler metric. We find that the model is well defined and gives a framework for describing anisotropic surface shape. More precisely, the FG surface model is obtained from the model of Hefrich and Polyakov (HP) for strings and membranes by replacing the Riemannian metric $g_{ab}$ with a Finsler metric $g_{ab}^{\rm F}$. In this sense, this model is an extension of the HP model. By discretizing the continuous FG model, we obtain a discrete FG model. The discrete Finsler length depends on the direction of local coordinates on the surface, where a new fictitious variable ${\mathcal C}$ is introduced. The variable ${\mathcal C}$ represents a coordinate on the triangles. 

We have confirmed that the model in this paper can be used as an anisotropic model for membranes step by step. Firstly, we study the most simple model with only fictitious variable  ${\mathcal C}$ in order to check that the treatment for the FG model (Euclid metric) is well defined. In this FG model, the crumpling transition between the crumpled and smooth phases is found to be of first-order and is identical to that of the conventional model. Secondly, we find that a FG surface model (Regge metric) with random vector field ${\bf v}$ has the same phase structure as the conventional model and confirm that the FG model is well defined. Thirdly, we define $g_{ab}^{\rm F}$ by using the Euclid metric and a constant vector field ${\bf v}$ in order to demonstrate that the model has an anisotropic phase. The Monte Carlo results show that a tubular surface is obtained in a certain range of the bending rigidity $\kappa$. Finally, we study an anisotropic model in which the tilt variable $\sigma$ is assumed to be dynamical. The in-plane components of $\sigma$ are used as the vector field to define the Finsler metric, and consequently the surface tension coefficient $\gamma$ and the bending rigidity $\kappa$ become dependent on the position and direction on the surface. We find three phases in the model; the spherical phase, the disk phase, and the tubular phase. 

Finally, some additional and speculative comments on the FG model are given. First of all, the advantage of the FG model over fixing $\gamma$ and $\kappa$ to be anisotropic by hand is that these quantities become dependent on the position and direction on the surface. This implies that the FG model is inhomogeneous in the sense that the strength of the surface force is not always uniform on the surface. Such inhomogeneity is well-known in biological membranes because of their internal structures like cytoskeletons or microtubules \cite{Kusumi-BioJ-2004,Hotani-AdvBioPhys-1990}. Because the microtubule structure is dynamically changeable due to the polymerization/depolymerization, the surface anisotropy caused by them is also expected to share a common property with the one caused by the molecular orientation property of the liquid crystal molecules in liquid crystal elastomer membranes. It is also possible to consider that a microscopic origin of Finsler metric is connected with an anisotropic random walk of molecules recalling that the ideal chain model for polymers is represented as an isotropic random walk \cite{Doi-Edwards-1986}.  
As mentioned in the previous section, further studies are needed to obtain more detailed information of the phase structure of Finsler geometric surface model.

%----------------------------------------------------------
\vspace*{3mm}
\noindent
{\bf Acknowledgments}\\
The authors thank Hiroki Mizuno for his support of computer analyses. The author (HK) acknowledges Andrey Shobukhov for discussions and comments, he is also grateful to Alexander Razgulin for discussions during a visit to Lomonosov Moscow State University, and he would like to thank Takashi Matsuhisa for discussions. This work is supported in part by a Promotion of Joint Research of Toyohashi University of Technology.

%\vspace*{2mm}

%\vfill\eject
%\vspace*{5mm}
%\noindent
%\section*{References}

\end{document}